\documentclass[aps,prc,preprint,amsmath,amssymb,showpacs,preprintnumbers,superscriptaddress]{revtex4-2}
\usepackage{CJK}
\usepackage{graphicx}
\usepackage{dcolumn}
\usepackage{bm}
\usepackage{color}
\usepackage{booktabs}
\usepackage{hyperref}

\allowdisplaybreaks[4]

\begin{document}

\title{Charge radii of calcium isotopes within relativistic configuration-interaction density functional theory}
\author{T. Qu}
\affiliation{State Key Laboratory of Nuclear Physics and Technology, School of Physics, Peking University, Beijing 100871, China}

\author{B. Li}
\affiliation{State Key Laboratory of Nuclear Physics and Technology, School of Physics, Peking University, Beijing 100871, China}

\author{Y. K. Wang}
\email{wangyk@buaa.edu.cn }
\affiliation{School of Physics, Beihang University, Beijing 102206, China}

\begin{abstract}
The charge radii of calcium isotopes are investigated within the framework of relativistic configuration-interaction density functional (ReCD) theory. 
The ReCD theory microscopically incorporates beyond-mean-field correlations through rotational symmetry restoration and configuration mixing among quasiparticle excited states, and treats even-even and odd-$A$ isotopes on the same footing. 
It is found that beyond-mean-field correlations significantly soften the potential energy surfaces of calcium isotopes and shift the energy minima from nearly spherical mean-field solutions to deformed shapes. 
The quadrupole deformation parameters predicted by the ReCD theory show much better agreement with the available experimental data than the mean-field results, supporting the reliability of the calculated potential energy surfaces and highlighting the important role of beyond-mean-field correlations.
Owing to the sensitive dependence of charge radii on nuclear deformation, the charge radii obtained within the ReCD framework are generally larger than the mean-field predictions. 
The nearly identical charge radii of $^{40}\mathrm{Ca}$ and $^{48}\mathrm{Ca}$, as well as the unexpectedly large charge radius of $^{52}\mathrm{Ca}$, are well reproduced. 
Compared with the mean-field calculations, the description of the odd-even staggering is improved, especially for the enhanced charge radii of $^{42}\mathrm{Ca}$ and $^{44}\mathrm{Ca}$. 
It is also worth noting that secondary local minima appear in the ReCD-based potential energy surfaces of the odd-$A$ calcium isotopes $^{41,43,47}\mathrm{Ca}$. 
The present results suggest that shape mixing between different local minima, which is not fully included in the present calculation, may further improve the description of the pronounced odd-even staggering observed in calcium isotopes.
\end{abstract}

\maketitle

\section{Introduction}~\label{sec.I}
Nuclear charge radius is one of the most important bulk properties of atomic nuclei. Its study is of great significance because the charge radius provides a sensitive probe of various nuclear structure phenomena, including shell evolution~\cite{Kreim2014PLB.731.97, Angeli2015NPG:NPP.42.055108, Gorges2019PRL.122.192502}, shape staggering~\cite{Marsh2018NP.14.1163, Barzakh2021PRL.127.192501}, shape coexistence~\cite{Yang2016PRL.116.182502}, neutron skin~\cite{Hagen2016NP.12.186, Yang2018PRC.97.014314, Gaidarov2020NPA.1004.122061}, and halo structures~\cite{Nortershauser2009PRL.102.062503}. Moreover, nuclear charge radii have been found to exhibit strong correlations with the saturation density and symmetry energy of nuclear matter~\cite{Reinhard2016PRC.93.051303, Brown2017PRL.119.122502, Pineda2021PRL.127.182503, Konig2024PRl.132.162502}, and therefore have been regarded as unique observables for constraining the nuclear equation of state and, consequently, the properties of neutron stars.
It is worth noting that subtle changes in nuclear structure are often manifested as sudden increases, decreases, or staggering patterns in the evolution of charge radii as functions of neutron or proton number. 
This makes systematic investigations along isotopic or isotonic chains a particularly powerful and widely used approach in nuclear physics. 
Owing to the rapid development of electron scattering~\cite{DeVries1987ADNDT.36.495}, muonic atom spectroscopy~\cite{Antwis2025SR.15.6939}, and collinear laser spectroscopy~\cite{Yang2023PPNP.129.104005}, an increasing amount of charge-radius data has become available over extended isotopic and isotonic chains; see the data compilation in Ref.~\cite{Angeli2013ADNDT.99.69} for more details.

Among the diverse isotopic trends revealed by charge-radius measurements, the calcium isotopes exhibit particularly intriguing behaviors~\cite{MartenssonPendrill1992PRA.45.4675, Vermeeren1992PRL.68.1679, Vermeeren1996JPG:NPP.22.1517, GarciaRuiz2016NP.12.594, Miller2019NP.15.432}. First, the charge radii of the two doubly magic nuclei $^{40}\mathrm{Ca}$ and $^{48}\mathrm{Ca}$ are found to be nearly identical, despite the addition of eight neutrons. Second, the charge radius of $^{52}\mathrm{Ca}$ is unexpectedly large, which is counterintuitive given the experimental evidence suggesting the doubly magic character of this nucleus~\cite{Wienholtz2013Nature.498.346}. 
Third, a pronounced odd-even staggering (OES) is observed in the charge radii of calcium isotopes between $^{36}\mathrm{Ca}$ and $^{48}\mathrm{Ca}$. 
These distinctive features have stimulated extensive theoretical investigations~\cite{Wang2013PRC.88.011301,Caurier2001PLB.522.240,Nakada2015PRC.92.044307,GarciaRuiz2016NP.12.594,Soma2020PRC.101.014318,Heinz2025PRC.111.034311,Reinhard2017PRC.95.064328,Perera2021PRC.104.064313}.

Most studies aimed at understanding the behavior of charge radii in calcium isotopes are carried out within the framework of density functional theory (DFT), and several extensions have been proposed to improve the theoretical description. 
Within nonrelativistic DFT, for instance, a density-dependent spin-orbit interaction was introduced in Ref.~\cite{Nakada2015PRC.92.044307}. 
This modification succeeds in reproducing the nearly identical charge radii of $^{48}\mathrm{Ca}$ and $^{40}\mathrm{Ca}$, but fails to capture the pronounced OES observed between $^{36}\mathrm{Ca}$ and $^{48}\mathrm{Ca}$.
In a different attempt~\cite{Reinhard2017PRC.95.064328}, gradient terms were incorporated into both the surface energy and pairing channel, leading to a reasonable reproduction of the isotopic evolution of charge radii after refitting the relevant parameters to selected charge-radii differential data. 
More recently, within relativistic DFT frameworks, phenomenological corrections have also been added directly to the calculated charge radii in Refs.\cite{An2020PRC.102.024307,An2024PRC.109.064302}, yielding an overall satisfactory agreement with experimental data.
Despite these successes, important limitations remain. 
In particular, the microscopic origin of the gradient terms introduced in Ref.~\cite{Reinhard2017PRC.95.064328} as well as the phenomenological corrections employed in Refs.~\cite{An2020PRC.102.024307,An2024PRC.109.064302} is unclear. 
Moreover, these approaches inevitably introduce additional parameters that must be constrained by selected charge-radius data in calcium isotopes.

The aforementioned extensions are all formulated at the mean-field level and exhibit certain limitations.
This naturally raises the question of how beyond-mean-field correlations affect the charge radii of calcium isotopes.
From a physical perspective, beyond-mean-field correlations are expected to play roles in nuclear charge radii.
In particular, such correlations give rise to dynamical shape fluctuations and trigger the development of nuclear dynamical deformation beyond the mean-field picture.
This, in turn, results in modifications of charge radii, as demonstrated in Bohr’s collective model~\cite{Bohr1975book}.
From an experimental viewpoint, the observation of the low-lying superdeformed band in $^{40}\mathrm{Ca}$ suggests that the potential energy surfaces of calcium isotopes may be significantly softer than those predicted by static mean-field calculations~\cite{Ideguchi2001PRL.87.222501}. 
This softness provides further evidence that beyond-mean-field effects should be properly taken into account in a quantitative description of charge radii.
Indeed, recent studies based on relativistic DFT (RDFT) combined with the five-dimensional collective Hamiltonian approach~\cite{Xie2025PRC.112.L021303}, which incorporates rotational and vibrational correlations induced by zero-point motion on top of the static mean-field solutions, have successfully reproduced not only the nearly identical charge radii of $^{40}\mathrm{Ca}$ and $^{48}\mathrm{Ca}$ but also the inverted parabolic trend between them. 
However, it is worth noting that this study was restricted to even-even calcium isotopes, and the impact of beyond-mean-field correlations on the OES of charge radii in calcium isotopes remains an open question.

In the present work, the charge radii of $^{39-52}\mathrm{Ca}$ are investigated within the framework of relativistic configuration-interaction density functional (ReCD) theory~\cite{Zhao2016PRC.94.041301,Wang2022PRC.105.054311}, a configuration-interaction approach built on microscopic RDFT.   
In the ReCD theory, beyond-mean-field correlations are incorporated through rotational symmetry restoration and configuration mixing among various quasiparticle excited states. 
These two ingredients provide a microscopic treatment of rotational and vibrational correlations associated with zero-point motion, respectively, in close connection with the beyond-mean-field effects considered in Ref.~\cite{Xie2025PRC.112.L021303}.
To date, ReCD theory has been successfully applied to the study of high-spin states~\cite{Zhao2016PRC.94.041301,Wang2022PRC.105.054311}, nuclear weak decays~\cite{Wang2024ScienceBulletin.69.2017, Wang2024PLB.855.138796}, nuclear chiral rotation~\cite{Wang2024PLB.848.138346}, and nuclear wobbling motion~\cite{Qu2025PRC.111.064309}, demonstrating strong predictive power in a variety of nuclear structure and decay phenomena. 
An important advantage of this framework is that even-even, odd-$A$, and odd-odd nuclei can be treated on the same footing. 
This feature makes ReCD theory particularly suitable for exploring how beyond-mean-field correlations influence the odd-even staggering observed in the charge radii of calcium isotopes.

This paper is organized as follows. 
In Sec.~\ref{sec.II}, the theoretical framework of ReCD theory is introduced. 
The numerical details are presented in Sec.~\ref{sec.III}. 
In Sec.~\ref{sec.IV}, the results are discussed. 
Finally, a summary and outlook are given in Sec.~\ref{sec.V}.

\section{Theoretical framework}~\label{sec.II}
The nuclear charge radius is evaluated as~\cite{Meng2006PPNP.57.470}
 \begin{equation}\label{chargeradius}
	r_\mathrm{c}=\sqrt{\langle \widehat{r^2_\mathrm{p}}\rangle+0.64}\quad(\mathrm{fm}).
\end{equation}
Here, the constant $0.64$ accounts for the finite size of the proton~\cite{Niksic2014CPC.185.1808}, and $\sqrt{\langle \widehat{r^2_\mathrm{p}}\rangle}$ denotes the root-mean-square radius of the point-proton density distribution. 
The corresponding mean-square radius is defined as
\begin{equation}\label{protonradius}
        \langle \widehat{r^2_\mathrm{p}}\rangle=\frac{1}{Z} \langle\Psi_{I}|\widehat{r^2_\mathrm{p}}|\Psi_{I}\rangle 
        = \frac{\langle\Psi_{I}|\widehat{r^2_\mathrm{p}}|\Psi_{I}\rangle}{\langle\Psi_{I}|\hat{N}_\mathrm{p}|\Psi_{I}\rangle},
\end{equation}
where $\widehat{r^2_\mathrm{p}}$ is the proton mean-square radius operator and $\hat{N}_\mathrm{p}$ is the proton-number operator.

The nuclear many-body wave function $|\Psi_I\rangle$ is obtained within the framework of ReCD theory and reads~\cite{Wang2024PLB.848.138346, Wang2024ScienceBulletin.69.2017,Wang2024PLB.855.138796}
\begin{equation}\label{wavefunction}
	|\Psi_\sigma^{IM}\rangle=\sum_{K\kappa}F^{I\sigma}_{K\kappa}\hat{P}^I_{MK}|\Phi_\kappa\rangle,
\end{equation}
where $\hat{P}^I_{MK}$ is the three-dimensional angular-momentum projection operator~\cite{Ring2004book}, and $\hat{P}^I_{MK}|\Phi_\kappa\rangle$ denotes the projected basis.
The state $|\Phi_\kappa\rangle$ refers to an intrinsic state in the configuration space. 
For even-even nuclei, the configuration space is constructed as
\begin{equation}\label{confspace-ee}
	\left\lbrace |\Phi_0\rangle,  \hat{\beta}^\dagger_{\pi_i}\hat{\beta}^\dagger_{\pi_j}|\Phi_0\rangle,  \hat{\beta}^\dagger_{\nu_i}\hat{\beta}^\dagger_{\nu_j}|\Phi_0\rangle,  \hat{\beta}^\dagger_{\pi_i}\hat{\beta}^\dagger_{\pi_j}\hat{\beta}^\dagger_{\pi_k}\hat{\beta}^\dagger_{\pi_l}|\Phi_0\rangle, \hat{\beta}^\dagger_{\nu_i}\hat{\beta}^\dagger_{\nu_j}\hat{\beta}^\dagger_{\nu_k}\hat{\beta}^\dagger_{\nu_l}|\Phi_0\rangle,  \hat{\beta}^\dagger_{\pi_i}\hat{\beta}^\dagger_{\pi_j}\hat{\beta}^\dagger_{\nu_k}\hat{\beta}^\dagger_{\nu_l}|\Phi_0\rangle\right\rbrace,
\end{equation}
whereas for odd-$N$ and even-$Z$ nuclei, it is given by
\begin{equation}\label{confspace-oa}
	\left\lbrace \hat{\beta}^\dagger_{\nu_0}|\Phi_0\rangle, \hat{\beta}^\dagger_{\nu_i}|\Phi_0\rangle,\hat{\beta}^\dagger_{\nu_i}\hat{\beta}^\dagger_{\pi_k}\hat{\beta}^\dagger_{\pi_l}|\Phi_0\rangle, \hat{\beta}^\dagger_{\nu_i}\hat{\beta}^\dagger_{\nu_k}\hat{\beta}^\dagger_{\nu_l}|\Phi_0\rangle\right\rbrace.
\end{equation}
Here, $\hat{\beta}^\dagger_\pi$ and $\hat{\beta}^\dagger_{\nu}$ represent the proton and neutron quasiparticle (qp) creation operators, respectively.
All intrinsic states in Eqs.~\eqref{confspace-ee} and \eqref{confspace-oa} are obtained by self-consistently solving the triaxial relativistic Hartree-Bogoliubov (TRHB) equation~\cite{Meng2016book}
\begin{equation}\label{RHB-equ}
	\begin{pmatrix}
		\hat{h}_D-\lambda&\hat{\Delta}\\
		-\hat{\Delta}^*&-\hat{h}_D^*+\lambda
	\end{pmatrix}
	\begin{pmatrix}
		U_k\\ V_k
	\end{pmatrix}
	=E_k
	\begin{pmatrix}
		U_k\\ V_k
	\end{pmatrix},
\end{equation}
where $U_k$ and $V_k$ are the qp wavefunctions, $\lambda$ is the Fermi energy, $\hat{h}_D$ is the single-particle Dirac Hamiltonian, and $\hat{\Delta}$ is the pairing field. 
For odd-$N$ and even-$Z$ nuclei, the quasineutron orbital $\nu_0$ with the lowest qp energy is blocked during the self-consistent iteration of the TRHB equation in order to ensure the correct number parity~\cite{Ring2004book}.

The weight coefficients $F^{I\sigma}_{K\kappa}$ in Eq.~\eqref{wavefunction} are determined by the following Hill-Wheeler equation
\begin{equation}\label{Hill-Wheeler}
	\sum_{K\kappa}\{H^{\prime I}_{K'\kappa' K\kappa} -E^{I\sigma}N^I_{K'\kappa'K\kappa}\}F^{I\sigma}_{K\kappa}=0,
\end{equation}
where $H^{\prime I}_{K'\kappa'K\kappa} = \langle\Phi_{\kappa'}|\hat{H}'\hat{P}^I_{K'K}|\Phi_\kappa\rangle$ and $N^I_{K'\kappa'K\kappa} = \langle\Phi_{\kappa'}|\hat{P}^I_{K'K}|\Phi_\kappa\rangle$ are respectively the energy kernel and the norm matrix.
These kernels can be evaluated by the Pfaffian algorithms proposed in Refs.~\cite{Carlsson2021PRL.126.172501, Hu2014PLB.734.162}.
The Hamiltonian $\hat{H}'$ is written as
\begin{equation}\label{Hamiltonian_prime}
	\hat{H}'=\hat{H}-\lambda_\mathrm{p}(\hat{N}_\mathrm{p}-Z)-\lambda_\mathrm{n}(\hat{N}_\mathrm{n}-N).
\end{equation}
Here, $\hat{H}=\hat{H}^\mathrm{RDFT}+\hat{H}^\mathrm{pair}$ is the Hamiltonian of the system. 
Since the intrinsic states $|\Phi_\kappa\rangle$ are not eigenstates of the proton and neutron number operators, the ReCD many-body wavefunctions are generally not eigenstates of the nucleon number operators either. 
To restore the correct mean values of the nucleon numbers, we follow the standard prescription~\cite{Bonche1990NPA.510.466,Hara1982NPA.385.14} and introduce the additional constraint terms $-\lambda_\mathrm{p}(\hat{N}_\mathrm{p}-Z)-\lambda_\mathrm{n}(\hat{N}_\mathrm{n}-N)$ in Eq.~\eqref{Hamiltonian_prime}.

The $\hat{H}^\mathrm{RDFT}$ can be derived from a relativistic Lagrangian density through a Legendre transformation~\cite{Meng2016book},
\begin{equation}\label{Hamiltonian}
	\begin{aligned}
		\hat{H}^\mathrm{RDFT}&=\int\mathrm{d}\bm{x} \Big\{\bar{\psi}(-\mathrm{i}\bm{\gamma}\cdot\nabla+M)\psi \\
		&+\frac{1}{2}\alpha_S(\bar{\psi}\psi)(\bar{\psi}\psi)+\frac{1}{2}\alpha_V(\bar{\psi}\gamma_\mu\psi)(\bar{\psi}\gamma^\mu\psi)+\frac{1}{2}\alpha_{TV}(\bar{\psi}\vec{\tau}\gamma_\mu\psi)\cdot(\bar{\psi}\vec{\tau}\gamma^\mu\psi)\\
		&-\frac{1}{2}\delta_S\partial_0(\bar{\psi}\psi)\partial^0(\bar{\psi}\psi)-\frac{1}{2}\delta_V\partial_0(\bar{\psi}\gamma_\mu\psi)\partial^0(\bar{\psi}\gamma^\mu\psi)-\frac{1}{2}\delta_{TV}\partial_0(\bar{\psi}\vec{\tau}\gamma_\mu\psi)\cdot\partial^0(\bar{\psi}\vec{\tau}\gamma^\mu\psi)\\
		&-\frac{1}{2}\delta_S\partial_i(\bar{\psi}\psi)\partial^i(\bar{\psi}\psi) -\frac{1}{2}\delta_V\partial_i(\bar{\psi}\gamma_\mu\psi)\partial^i(\bar{\psi}\gamma^\mu\psi)-\frac{1}{2}\delta_{TV}\partial_i(\bar{\psi}\vec{\tau}\gamma_\mu\psi)\cdot\partial^i(\bar{\psi}\vec{\tau}\gamma^\mu\psi)\\
		&	-\frac{1}{2}F^{0\nu}\dot{A}_\nu+\frac{1}{2}F^{i\nu}\partial_i A_\nu+e\bar{\psi}\frac{1-\tau_3}{2}\gamma^\mu A_\mu\psi\Big\}.
	\end{aligned}
\end{equation}
Pairing correlations are treated by the pairing Hamiltonian
\begin{equation}\label{Hpair}
	\hat{H}^\mathrm{pair}=\frac{1}{4}\sum_{\alpha\beta\gamma\delta}\langle\alpha\beta|\hat{V}^\mathrm{pair}|\gamma\delta\rangle\hat{c}^\dagger_\alpha\hat{c}^\dagger_\beta\hat{c}_\delta\hat{c}_\gamma.
\end{equation}
For $\hat{V}^\mathrm{pair}$, we adopt a  finite-range separable force~\cite{Tian2009PLB.676.44},
\begin{equation}\label{Vpair}
	\hat{V}^\mathrm{pair}=G\delta(\bm{R}-\bm{R}')P(\bm{r})P(\bm{r'})\frac{1}{2}(1-P^\sigma).
\end{equation}
Here, $\bm{R}=(\bm{r}_1+\bm{r}_2)/2$ and $\bm{r}=\bm{r}_1-\bm{r}_2$ are the center-of-mass and relative coordinates, respectively, and $(1-P^\sigma)/2$ projects onto the $S=0$ spin-singlet channel. 
More details on the separable pairing force can be found in Ref.~\cite{Tian2009PLB.676.44}.
Combining Eqs.~\eqref{protonradius} and \eqref{wavefunction}, one obtains the final expression for the point-proton mean-square radius within the framework of ReCD theory,
\begin{equation}\label{protonradius-v1}
	\langle \widehat{r^2_\mathrm{p}}\rangle= \frac{\sum\limits_{K K'\kappa\kappa'}F^{I\sigma*}_{K'\kappa'}F^{I\sigma}_{K\kappa}\langle\Phi_{\kappa'}|\widehat{r^2_\mathrm{p}}\hat{P}^{I}_{K'K}|\Phi_\kappa \rangle}{\sum\limits_{K K'\kappa\kappa'}F^{I\sigma*}_{K'\kappa'}F^{I\sigma}_{K\kappa}\langle\Phi_{\kappa'}|\hat{N}_\mathrm{p}\hat{P}^{I}_{K'K}|\Phi_\kappa \rangle}.
\end{equation}

\section{Numerical details}~\label{sec.III}
In the present work, the charge radii of $^{39-52}\mathrm{Ca}$ are studied within the framework of ReCD theory. 
Both the Hamiltonian $\hat{H}^\mathrm{RDFT}$ and the TRHB equation are derived from the relativistic density functional PC-PK1~\cite{Zhao2010PRC.82.054319}. For the pairing Hamiltonian $\hat{H}^\mathrm{pair}$, the standard pairing strength $G = 728~\mathrm{MeV \cdotp fm^3}$ suggested in Ref.~\cite{Tian2009PLB.676.44} is adopted. 
The TRHB equation is solved in a three-dimensional harmonic-oscillator basis with $10$ major shells.
Since the present study focuses on the charge radii of ground states, the contributions from the four-qp configurations in Eq.\eqref{confspace-ee} and the three-qp configurations in Eq.\eqref{confspace-oa} are expected to be negligible. 
Therefore, when constructing the configuration space, only two-qp intrinsic states are included for even-even calcium isotopes, and one-qp intrinsic states are included for odd-$A$ isotopes.
Similar to our previous studies~\cite{Zhao2016PRC.94.041301,Wang2022PRC.105.054311,Wang2024PLB.848.138346,Wang2024PLB.855.138796, Wang2024ScienceBulletin.69.2017,Qu2025PRC.111.064309}, a qp excitation-energy cutoff $E_\mathrm{cut}$ is introduced to truncate the dimension of the configuration space. 
In the present calculations, $E_\mathrm{cut}=3.5~\mathrm{MeV}$ is adopted, which ensures the convergence of the relevant results discussed below.

\section{Results and discussion}~\label{sec.IV}
\begin{figure}[!htbp]
  \centering
  \includegraphics[width=0.9\textwidth]{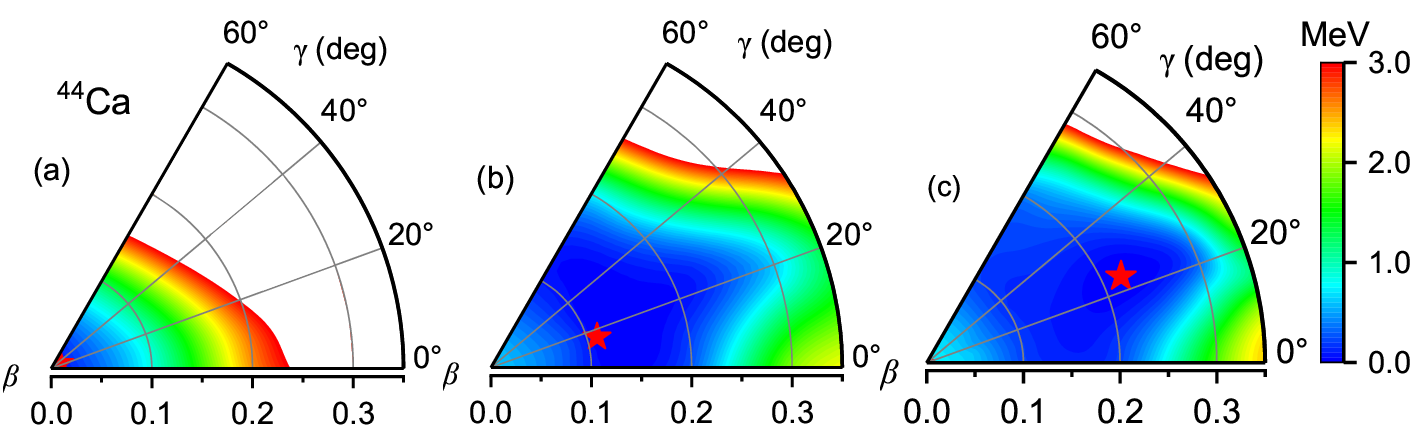}
  \caption{(Color online) Potential energy surfaces (PESs) of $^{44}\mathrm{Ca}$ in the plane of the axially symmetric quadrupole deformation parameter $\beta$ and the triaxial deformation parameter $\gamma$.
(a) PES of the mean-field states obtained by solving the TRHB equation.
(b) PES of the $0^+$ states obtained after angular-momentum projection of the mean-field states.
(c) PES of the $0^+$ states obtained after angular-momentum projection and configuration mixing, corresponding to the full ReCD calculation.}
  \label{Fig:PES44Ca}
\end{figure}

As discussed above, beyond-mean-field correlations can induce dynamical shape fluctuations and drive the nucleus toward equilibrium deformations that differ from those predicted at the mean-field level. 
To illustrate this effect, we take $^{44}\mathrm{Ca}$ as an example and present in Fig.~\ref{Fig:PES44Ca} the potential energy surfaces (PESs) obtained from the TRHB mean-field calculations, from angular-momentum-projected states built on top of the mean-field solutions, and from the full ReCD calculation including both angular momentum projection and configuration mixing.
At the mean-field level, the PES exhibits its minimum at $\beta=0$, indicating that $^{44}\mathrm{Ca}$ is predicted to be spherical. 
Moreover, the surface is relatively stiff, with the excitation energy increasing rapidly up to about $2.5~\mathrm{MeV}$ at $\beta=0.2$. 
After angular momentum projection is performed, the minimum of the corresponding PES shifts to $(\beta,\gamma) = (0.11,16^\circ)$, and the surface becomes significantly softer compared to the mean-field result, as shown in Fig.~\ref{Fig:PES44Ca} (b).
Such a soft PES implies that additional correlation effects can further modify the equilibrium shape. 
Indeed, after including additional correlations induced by qp configuration mixing within the full ReCD framework, the minimum of the PES is further shifted to a larger deformation, located at $(\beta,\gamma) = (0.21,25^\circ)$, as shown in Fig.~\ref{Fig:PES44Ca} (c). 
The resulting shape changes are expected to have a direct impact on the charge radii, as discussed in the following.

\begin{figure}[!htbp]
	\centering
	\includegraphics[width=0.9\textwidth]{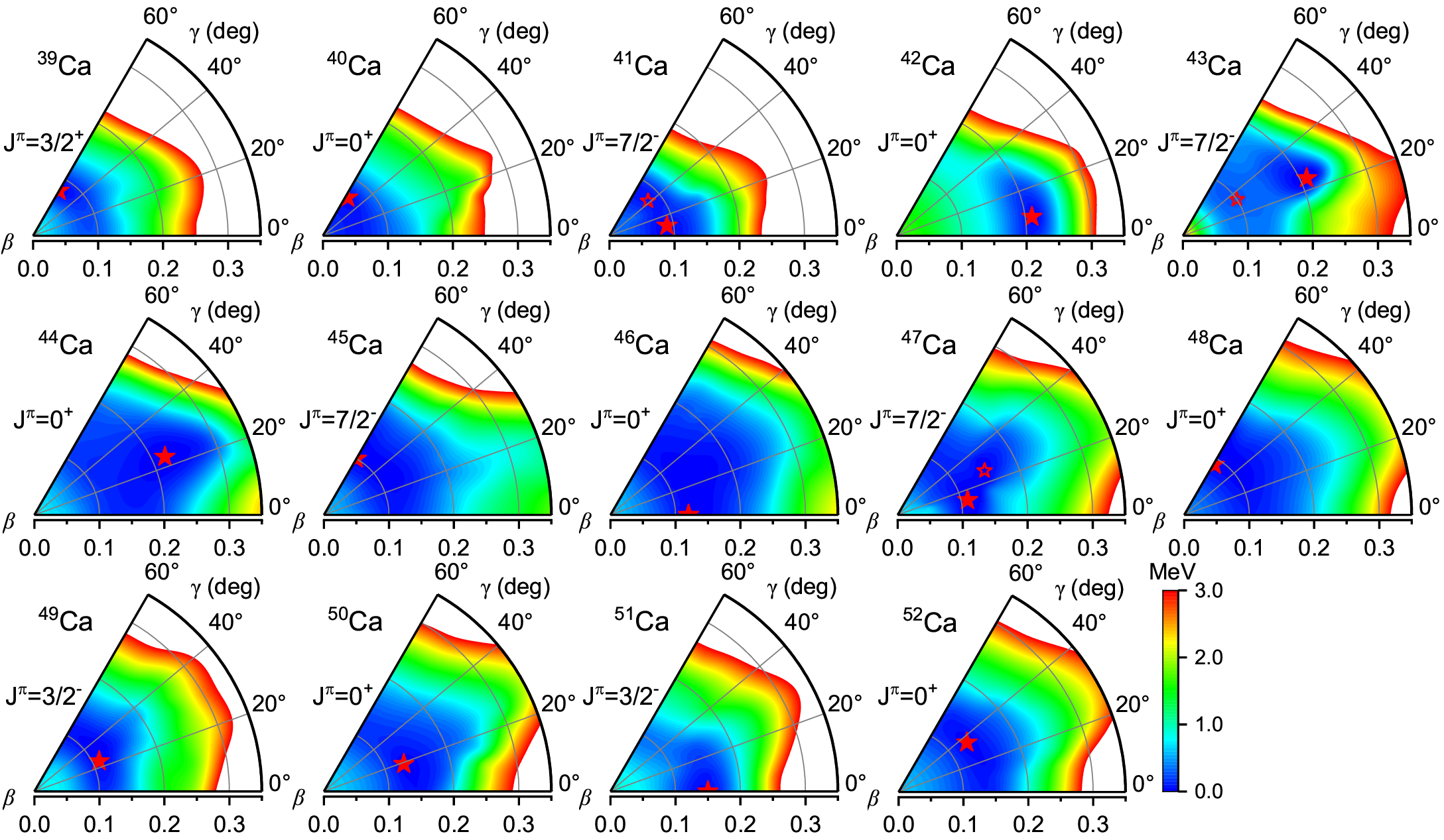}
	\caption{(Color online) Same as Fig.~\ref{Fig:PES44Ca}(c), but for the calcium isotopes from $^{39}\mathrm{Ca}$ to $^{52}\mathrm{Ca}$.
	The red solid stars indicate the positions of the global minima, while the red open stars denote the corresponding secondary local minima.}
	\label{Fig:PES}
\end{figure}

It is worth noting that the shape evolution induced by beyond-mean-field correlations appears to be a general feature across the calcium isotopic chain. 
In Fig.~\ref{Fig:PES}, the complete PESs for calcium isotopes from $^{39}\mathrm{Ca}$ to $^{52}\mathrm{Ca}$ obtained within the ReCD framework are displayed. 
It is encouraging to see that the ReCD theory correctly reproduces the spin and parity $J^\pi$ of odd-$N$ calcium isotopes, which provides a positive indication of the predictive power of the present approach.
A salient feature of the ReCD-based PESs is the pronounced softness along the entire $^{39-52}\mathrm{Ca}$ isotopic chain, in clear contrast to the relatively stiff surfaces predicted by mean-field calculations (see, e.g., TRHB results in Ref.~\cite{Yang2020TRHB,Yang2021PRC.104.054312}). 
Moreover, the equilibrium shapes exhibit a rich variety across the isotopic chain. 
For $^{39,40,45,48}\mathrm{Ca}$, the energy minima are located at $\gamma = 60^\circ$, indicating oblate-dominated deformation. 
In contrast, $^{46}$Ca and $^{51}$Ca exhibit minima at $\gamma = 0^\circ$, corresponding to prolate-dominated shapes. 
Interestingly, for $^{41-44,47,49,50,52}\mathrm{Ca}$, the minima are found at intermediate values of $\gamma$, away from both $0^\circ$ and $60^\circ$, indicating the presence of triaxial deformation.
These equilibrium deformations differ significantly from those predicted by the TRHB mean-field calculations, where all calcium isotopes are found to be nearly spherical. 
This clear shift in the location of the energy minima highlights the essential role of beyond-mean-field correlations in shaping the structure of calcium isotopes.

\begin{figure}[!htbp]
	\centering
	\includegraphics[width=0.5\textwidth]{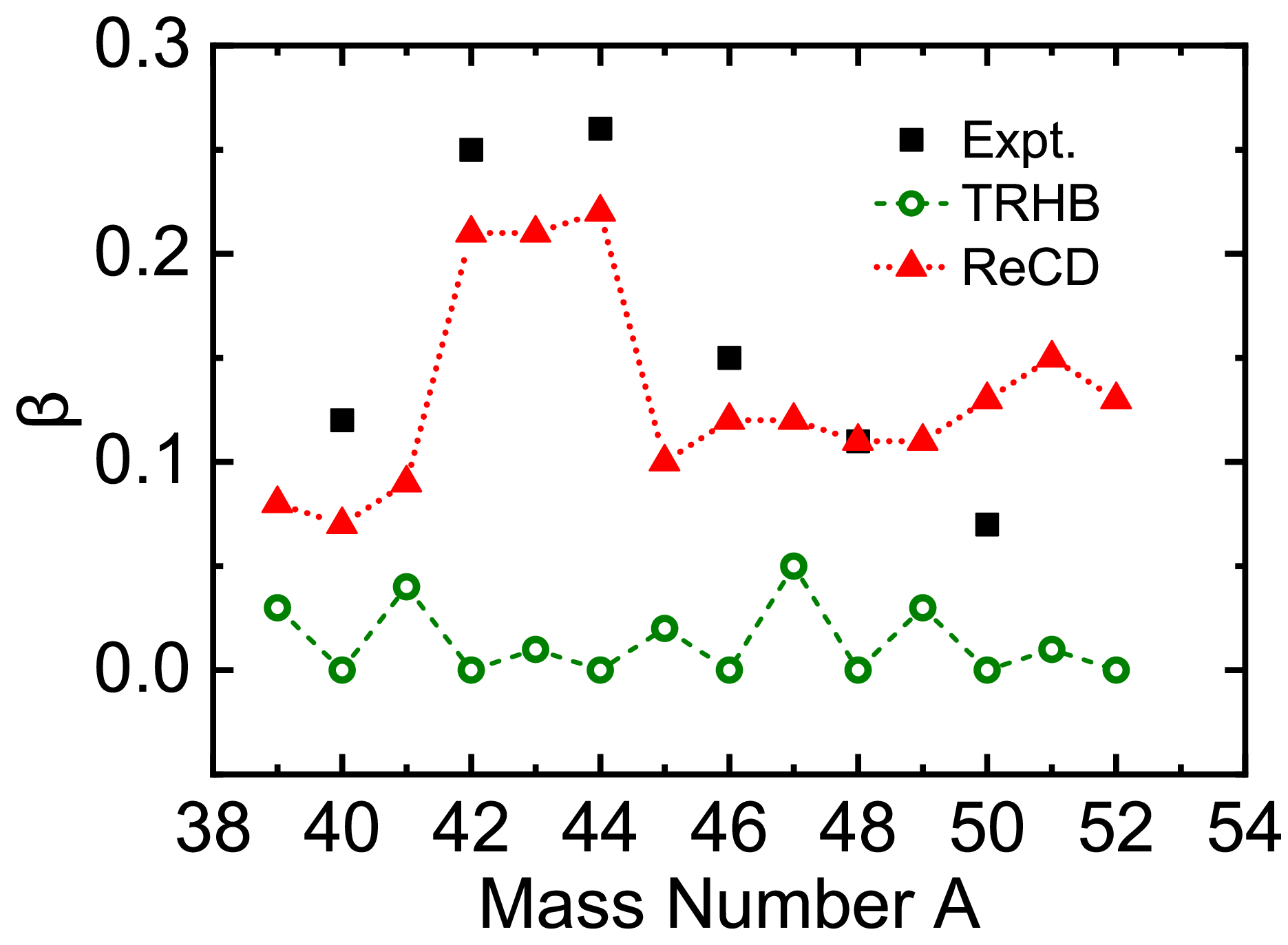}
	\caption{(Color online) The quadrupole deformation parameter $\beta$ at the positions of the energy minima on the PESs for $^{39-52}\mathrm{Ca}$, obtained from the ReCD calculations (red solid triangles), in comparison with the results from TRHB calculations (green open circles) and available experimental data (black solid squares)~\cite{NNDC2023}.}
	\label{Fig:beta2}
\end{figure}

The deformation shifts with respect to the mean-field results are physically expected. 
However, from a quantitative perspective, it remains necessary to examine whether the ReCD-predicted shapes are consistent with the available experimental data.
In Fig.~\ref{Fig:beta2}, the quadrupole deformation parameter $\beta$ at the minima of the PESs shown in Fig.~\ref{Fig:PES} is presented, together with a comparison to the TRHB results and the experimental values extracted from $B(E2)$ data~\cite{NNDC2023}. 
It is evident that the experimental $\beta$ values effectively include dynamical correlations that are absent in static mean-field calculations. 
As a consequence, all available experimental $\beta$ values for even-even calcium isotopes are nonzero, with particularly large values of $\beta \approx 0.25$ for $^{42}\mathrm{Ca}$ and $^{44}\mathrm{Ca}$, in stark contrast to the nearly spherical shapes predicted by the TRHB calculations.
When beyond-mean-field correlations are taken into account, the ReCD results reproduce the experimental $\beta$ values reasonably well. 
These agreements provide further support for the reliability of the PESs shown in Fig.~\ref{Fig:PES}, and once again highlight the essential role of beyond-mean-field correlations in describing the structural properties of calcium isotopes.

\begin{figure}[!htbp]
	\centering
	\includegraphics[width=0.5\textwidth]{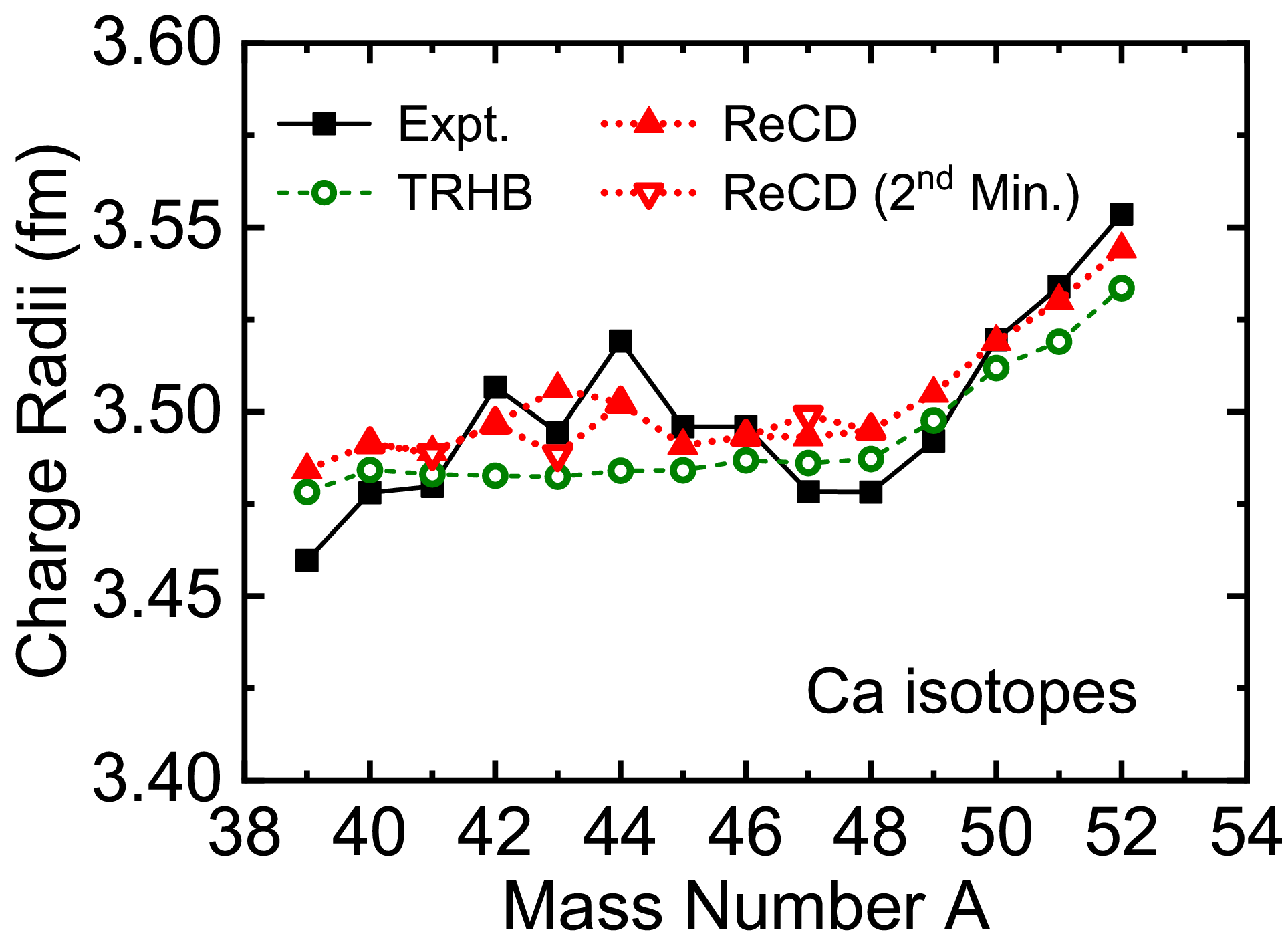}
	\caption{(Color online) Charge radii of calcium isotopes calculated within the ReCD framework (red solid triangles), in comparison with results from TRHB mean-field calculations (green open circles) and available experimental data (black solid squares)~\cite{GarciaRuiz2016NP.12.594}. 
	The red open triangles correspond to the charge radii obtained from the many-body wavefunctions associated with the secondary local minima shown in Fig.~\ref{Fig:PES}; see the text for details.
	}
	\label{Fig:chargeradii}
\end{figure}

It is reasonable to expect that deformation changes induced by beyond-mean-field correlations will influence the resulting charge radii, as suggested by Bohr’s collective model, $\langle r^2\rangle_{\text{c}} = \langle r^2\rangle_{\beta = 0} \left( 1 + \frac{5}{4\pi} \beta^2 \right)$.
To microscopically demonstrate the beyond-mean-field effects on charge radii, we present in Fig.~\ref{Fig:chargeradii} the charge radii obtained from the ReCD theory, in comparison with TRHB mean-field results and available experimental data.
As discussed previously, the intrinsic states $|\Phi_\kappa\rangle$ are not eigenstates of the proton-number operator, and consequently the ReCD many-body wave functions $|\Psi_I\rangle$ are not exact eigenstates of $\hat{N}_\mathrm{p}$. 
To mitigate the effects of particle-number nonconservation, the constraint terms $-\lambda_\mathrm{p}(\hat{N}_\mathrm{p}-Z)-\lambda_\mathrm{n}(\hat{N}_\mathrm{n}-N)$ are introduced, as shown in Eq.~\eqref{Hamiltonian_prime}. 
The validity of this prescription for the total nuclear energy has been well established~\cite{Bonche1990NPA.510.466,Hara1982NPA.385.14}. 
However, in practical calculations, we find that the expectation value $\langle \Psi_I|\hat{N}_\mathrm{p}|\Psi_I\rangle$ slightly deviates from the exact proton number $Z=20$, and this deviation becomes more pronounced when quasiparticle configuration mixing is included. 
Since the evaluation of charge radii depends on properly normalized proton densities [see Eq.~\eqref{protonradius-v1}], such deviations may affect the calculated radii.
To avoid this contamination, we neglect the contributions from excited quasiparticle configurations when evaluating charge radii, i.e., we omit the summation over the index $\kappa$ in Eq.~\eqref{protonradius-v1}. 
This is an approximation, and a more rigorous treatment would require an explicit particle-number projection of the ReCD many-body wavefunctions, which will be pursued in future work.

As shown in Fig.~\ref{Fig:chargeradii}, the TRHB mean-field calculations reproduce the nearly identical charge radii of $^{40}\mathrm{Ca}$ and $^{48}\mathrm{Ca}$, as well as the anomalously large charge radius of the neutron-rich isotope $^{52}\mathrm{Ca}$. 
However, they significantly underestimate the charge radii in the region from $^{42}\mathrm{Ca}$ to $^{46}\mathrm{Ca}$, and fail to reproduce the observed OES.
Compared with the TRHB results, the ReCD calculations generally yield larger charge radii. 
In particular, both the near degeneracy of the charge radii of $^{40}\mathrm{Ca}$ and $^{48}\mathrm{Ca}$ and the enhanced radius of $^{52}\mathrm{Ca}$ are well reproduced. 
More importantly, the pronounced OES between $^{41}\mathrm{Ca}$ and $^{42}\mathrm{Ca}$, as well as between $^{44}\mathrm{Ca}$ and $^{45}\mathrm{Ca}$, is also well described.
Nevertheless, the kink observed at $^{43}\mathrm{Ca}$ is not reproduced within the present ReCD framework, suggesting that the predicted deformation may be too large and that additional correlations are still missing. 
It is noteworthy that a secondary minimum appears on the PES of $^{43}\mathrm{Ca}$ (see Fig.~\ref{Fig:PES}). 
Interestingly, the charge radius associated with this secondary minimum is in better agreement with the experimental data.
In fact, as shown in Fig.~\ref{Fig:PES}, the PESs of calcium isotopes are generally soft, indicating that large-amplitude shape fluctuations may play an important role. 
Such effects can, in principle, only be properly described by approaches such as the generator coordinate method or the five-dimensional collective Hamiltonian method. 
Incorporating large-amplitude shape fluctuations by mixing different nuclear shapes within the ReCD framework, and exploring their impact on the charge radii of calcium isotopes, therefore represents an important direction for future work.

\section{Summary and outlook}~\label{sec.V}
In summary, the charge radii of $^{39-52}\mathrm{Ca}$ have been investigated within the framework of relativistic configuration-interaction density functional (ReCD) theory, which incorporates beyond-mean-field correlations through rotational symmetry restoration and configuration mixing, and treats even-even and odd-$A$ isotopes on the same footing. 
It is found that beyond-mean-field correlations soften the potential energy surfaces and shift the equilibrium shapes of calcium isotopes from the nearly spherical mean-field solutions to deformed configurations. 
Taking $^{44}\mathrm{Ca}$ as an example, we have illustrated in detail how the potential energy surface evolves when the beyond-mean-field correlations induced by rotational symmetry restoration and configuration mixing are successively taken into account.
The reliability of the calculated potential energy surfaces and the corresponding deformation parameters is supported by their satisfactory agreement with the experimental deformation values extracted from $B(E2)$ data. 
In particular, the large deformations of $^{42}\mathrm{Ca}$ and $^{44}\mathrm{Ca}$, which are predicted to be nearly spherical by the mean-field calculations, are well reproduced within the ReCD framework. 
Owing to the sensitive dependence of charge radii on nuclear deformation, the charge radii obtained after including beyond-mean-field correlations are generally larger than the mean-field predictions.
The nearly identical charge radii of $^{40}\mathrm{Ca}$ and $^{48}\mathrm{Ca}$, as well as the unexpectedly large charge radius of $^{52}\mathrm{Ca}$, are successfully reproduced. 
More importantly, the pronounced odd-even staggering between $^{41}\mathrm{Ca}$ and $^{42}\mathrm{Ca}$, as well as between $^{44}\mathrm{Ca}$ and $^{45}\mathrm{Ca}$, is also well described, mainly due to the enhanced deformations predicted for $^{42}\mathrm{Ca}$ and $^{44}\mathrm{Ca}$. 
Nevertheless, the odd-even staggering between $^{42}\mathrm{Ca}$ and $^{43}\mathrm{Ca}$ is not reproduced in the present calculations, as the charge radius of $^{43}\mathrm{Ca}$ is overestimated.
For $^{43}\mathrm{Ca}$, we find that a secondary local minimum, with an energy close to that of the global minimum, appears in the potential energy surface. 
The charge radius obtained from the many-body wavefunction associated with this secondary minimum shows better agreement with the experimental data. 
This finding highlights the possible importance of shape mixing, which is not fully captured in the present framework, for describing the charge radii of odd-$A$ calcium isotopes. 
Therefore, incorporating shape mixing, for example through a generator-coordinate-method extension of the present ReCD theory, and investigating its impact on the odd-even staggering of charge radii in calcium isotopes will be an interesting direction for future work.

\begin{acknowledgments}
This work has been supported in part by the National Key Research and Development Program of China (Grants No. 2024YFA1612600, No. 2024YFE0109803), National Natural Science Foundation of China (Grants No. 12141501, No. 12105004, No. 12475117, No. 12435006), the Beijing Natural Science Foundation under Grant No. 1242019, the National Key Laboratory of Neutron Science and Technology under Grant No. NST202401016, the China Postdoctoral Science Foundation under Grant No. 2020M680183, the State Key Laboratory of Nuclear Physics and Technology, Peking University (Grants No. NPT2023KFY05, NPT2023ZX03), and the High-performance Computing Platform of Peking University. 
\end{acknowledgments}

\end{document}